\documentclass[aps,prc,twocolumn,superscriptaddress]{revtex4-2}

\usepackage{graphicx}
\usepackage{dcolumn}
\usepackage[utf8]{inputenc}
\usepackage{comment}
\usepackage{multirow, tabularray}
\usepackage[table,xcdraw]{xcolor}
\usepackage{float}

\definecolor{color_reso1}{rgb}{0.027, 0.553, 0.43}
\definecolor{color_reso2}{rgb}{0.149, 0.808, 0.667}
\definecolor{color_reso345}{rgb}{0.596, 0.91, 0.757} 
\definecolor{color_reso6}{rgb}{0.482, 0.678, 0.886}
\definecolor{color_reso78}{rgb}{0.314, 0.286, 0.8} 
\definecolor{color_reso910}{HTML}{20A7C2}

\begin{document}

\preprint{APS/123-QED}

\title{Measurement of the \textsuperscript{35}Cl$(n, p)$\textsuperscript{35}S cross-section at the CERN n\_TOF facility \\ from subthermal energy to 120 keV}

\author{Marco Antonio Martinez-Cañadas}  \affiliation{University of Granada, Spain}%
\author{Pablo~Torres-S\'{a}nchez} \email{Pablo.Torres@ific.uv.es}  \affiliation{University of Granada, Spain} \affiliation{Instituto de F\'{\i}sica Corpuscular, CSIC - Universidad de Valencia, Spain} %
\author{Javier~Praena} \affiliation{University of Granada, Spain} %
\author{Ignacio~Porras} \affiliation{University of Granada, Spain} %
\author{Marta~Sabat\'{e}-Gilarte} \affiliation{European Organization for Nuclear Research (CERN), Switzerland} \affiliation{Universidad de Sevilla, Spain} %
\author{Oliver~Aberle} \affiliation{European Organization for Nuclear Research (CERN), Switzerland} %
\author{Victor~Alcayne} \affiliation{Centro de Investigaciones Energ\'{e}ticas Medioambientales y Tecnol\'{o}gicas (CIEMAT), Spain} %
\author{Simone~Amaducci} \affiliation{INFN Laboratori Nazionali del Sud, Catania, Italy} \affiliation{Dipartimento di Fisica e Astronomia, Universit\`{a} di Catania, Italy} %
\author{J\'{o}zef~Andrzejewski} \affiliation{University of Lodz, Poland} %
\author{Laurent~Audouin} \affiliation{Institut de Physique Nucl\'{e}aire, CNRS-IN2P3, Univ. Paris-Sud, Universit\'{e} Paris-Saclay, F-91406 Orsay Cedex, France} %
% ...
%

\collaboration{The n\_TOF Collaboration (www.cern.ch/ntof)} \noaffiliation

\date{\today}

\begin{abstract} 
\noindent \textbf{Background:} 
The \textsuperscript{35}Cl$(n, p)$\textsuperscript{35}S reaction is of special interest in three different applications. First, in Boron Neutron Capture Therapy due to the presence of \textsuperscript{35}Cl in brain and skin tissue. Second, it is involved in the creation of \textsuperscript{36}S, whose astrophysical origin remains unresolved. Third, in the designing of fast nuclear reactors of new generation based on molten salts. \\
\textbf{Purpose:} 
To measure the \textsuperscript{35}Cl$(n, p)$\textsuperscript{35}S cross-section from thermal energy to 120\,keV, determine the resonance parameters in this range and Maxwellian Averaged Cross-Section (MACS).\\
\textbf{Method:} 
We made use of the Time-of-Flight technique with microMEGAS detectors at Experimental Area 2 (EAR-2) of n\_TOF facility at CERN. The \textsuperscript{10}B$(n, \alpha)$\textsuperscript{7}Li and \textsuperscript{235}U$(n, f)$ reactions were used as references. Rutherford Back-scattering Spectrometry technique was performed at Centro Nacional de Aceleradores (CNA) in Sevilla, in order to accurately determine the masses of the irradiated samples. \\
\textbf{Results:} 
We obtain a thermal cross-section of $0.470 \pm 0.009$ barns. The $1/v$ energy dependence of the cross-section is observed up to the first resonance at 0.398\,keV, the resonances up to 120\,keV are analyzed and MACS calculated for $k_B T$ from 1 to 100\,keV. \\
\textbf{Conclusions:}
The  \textsuperscript{35}Cl$(n, p)$\textsuperscript{35}S cross-section has been obtained over a wide energy range for the first time, with high accuracy across the aforementioned range. The thermal cross-section and first two resonances are in agreement with latest evaluation in ENDF/B-VIII.1, while lower resonance strength was found for high energy resonances. These data are used to calculate the MACS for different $k_B T$.
\end{abstract}

\maketitle

\section{\label{sec:Introduction}Introduction}
\subsection{\label{sec:sec:Applications}Motivations}
There are several applications for the cross-section of \textsuperscript{35}Cl$(n, p)$\textsuperscript{35}S reaction: radiotherapy, astrophysics and nuclear energy production.

In Boron Neutron Capture Therapy (BNCT), the dose delivered in \textsuperscript{10}B-loaded tumor tissue due to the \textsuperscript{10}B$(n, \alpha)$\textsuperscript{7}Li reaction is more than four times higher than the contribution of all the others reactions \cite{Kankaanranta2011}, however, the limiting dose is given by the dose received in healthy tissue, where the reactions with other nuclei are more dominant. The International Commission on Radiation Units and Measurements (ICRU) recommends that the delivered dose should always have less than 5\% deviation from the prescribed one \cite{ICRU1992}. An accurate cross-section of the involved isotopes is one of the most important data to be taken into account in such calculations. Even though the concentration of \textsuperscript{35}Cl in brain tissue (one of the main indications of BNCT) is lower than that of other elements (0.2\%), its contribution to the dose is non-negligible, particularly due to the strong resonances at 0.397 and 4.25\,keV  \cite{Goorley2002}.

In stellar environments, the \textsuperscript{35}Cl$(n, p)$\textsuperscript{35}S reaction takes part in the synthesis of the rare isotope \textsuperscript{36}S, whose origin remains unresolved \cite{Schatz1995}. Recent observations of the circumstellar envelope of carbon-rich AGB stars show evidence of an important deviation from solar isotopic ratio of heavier elements than those involved in the Carbon-Nitrogen-Oxygen cycle, including (\textsuperscript{35}Cl/\textsuperscript{37}Cl) \cite{Kahane2000}. Current discrepancies in the \textsuperscript{35}Cl MACS \cite{Druyts1994} and gaps at certain stellar energies preclude from reliably determining the role of the \textsuperscript{35}Cl$(n, p)$\textsuperscript{35}S reaction on \textsuperscript{36}S abundances and in the aforementioned deviation of the solar isotopic ratio.

The designing of generation IV fast nuclear reactors considers the usage of molten chloride salts in order to use spent fuel from light-water reactors, which would reduce environmental consequences \cite{Tahara2024}. However, there is a large uncertainty in reactivity for such reactors mainly due to the uncertainty in the cross-section of \textsuperscript{35}Cl$(n, p)$\textsuperscript{35}S reaction in the resonances region and higher energies (0.4\,keV onwards), which propagates to other crucial safety parameters like reactivity coefficients. Production of \textsuperscript{35}S via this reaction is also important for corrosion. This cross-section is present in the NEA Nuclear Data High Priority Request List (HPRL) \cite{Dupont2020HPRL, NEA_HPRL_2022}. Any uncertainty reduction with respect to current data would be valuable, and our measurement covers the HPRL range of interest. Furthermore, our data at energies below resonances are very valuable for future evaluations of this reaction.

\subsection{\label{sec:sec:Previous_measurements}Previous measurements}
Several measurements have been performed at the thermal energy (25.3\,meV) \cite{GHESims,DJHughes,IGSchroeder,AGibert,HBerthet,Druyts1994,Gledenov1999}. Contrary to this, experimental data covering the resonance region is more scarce.

Specifically, Popov \textit{et al.} \cite{popov1961}  measured the cross-section in 1961 using a lead slowing-down spectrometer from 10 eV to 8\,keV with a poor resolution (from 35 to 70\%) and normalized their data to thermal cross-section of 0.19 b. They also reported a resonance at 1\,keV which has not been observed in any later experiment. 

Koehler \textit{et al.} \cite{Koehler1991}  measured the cross-section from 25\,meV to 100\,keV in 1991 at the LANSCE facility, and normalized their data to 0.489 b \cite{ATLAS_Mughabghab}. Data from Koehler \textit{et al.} had better energy resolution compared to Popov \textit{et al.}. However, later studies have suggested \cite{Druyts1994} that potential anisotropic proton emission  may have introduced systematic uncertainties due to the limited solid angle coverage in the setup.

Druyts \textit{et al.} \cite{Druyts1994} provided in 1994 a value of cross-section at thermal energy (measured at ILL) and restricted energy ranges around a few resonances (measured at GELINA). Gledenov \textit{et al.} \cite{Gledenov1999} later identified (in their own words) `the poor knowledge of the neutron flux' as a potential issue on the data from Druyts \textit{et al.}

As for Gledenov \textit{et al.} \cite{Gledenov1999}, the experiment was performed at the IBR-30 pulsed reactor of JINR in 1989 but remained unpublished until 1999. In practice, Gledenov \textit{et al.} themselves were aware of some technical difficulties present in their measurement. Specifically, they had doubts about the accuracy of the weighting technique that was performed and contamination from \textsuperscript{14}N(n,p) present in the gas filling of their ionization chamber. 

The evaluation in ENDF/B-VIII.1 (unchanged since release of ENDF/B-VII \cite{Chadwick20112887}, grounded on \cite{Sayer2006}) is based on three experimental data sets from Koehler \textit{et al.}, Druyts \textit{et al.} and Gledenov \textit{et al.} Some concerns were expressed in this evaluation: the resonances of Druyts \textit{et al.} could be fitted only when using the thermal value taken by Koehler \textit{et al.}: 0.489 b. This value was taken from \cite{GHESims} and after being re-calculated by \textit{et al.} considering more recent values of \textsuperscript{59}Co thermal cross-section and \textsuperscript{35}Cl abundance in natural chlorine, it becomes 0.483 b. However, it is still 12\% higher than the thermal value reported by Druyts \textit{et al.} (0.440 $\pm$ 0.010 b). The ENDF evaluation claimed that it was not possible to obtain a simultaneous fitting of all data with a thermal value significantly below 0.483 b. Further, there are unexplained significant discrepancies between resonance strengths of Druyts \textit{et al.} and Koehler \textit{et al.}, and consequently in the MACS; above 14\,keV, the evaluation \cite{Sayer2006} is based only on Druyts. Other databases such as JEFF or JENDL use the same evaluation as ENDF-VIII.1 below 1\,MeV.

In this work we provide a measurement avoiding the deficiencies listed by previous works, i.e., coverage of entire 4$\pi$ in angular distribution from two back-to-back samples, a proper knowledge of neutron flux, an accurate determination of the mass of the samples, and a good characterization and correction of the background.

\section{\label{sec:Introduction}Experimental setup}
\subsection{\label{sec:sec:nTOF}The n\_TOF facility}
The experiment was performed at the Experimental Area 2 (EAR-2) of the n\_TOF facility at CERN. EAR-2 is located about 19.5 meters away from the neutron production target in the vertical direction. Neutrons are generated by the 20 GeV/c proton pulse beam from the CERN Proton Synchrotron impinging onto a lead target. In-depth technical features of the facility and the characteristics of the neutron beam produced are described in detail in Refs. \cite{Weiss2015, Sabategilarte2017}.

\subsection{\label{sec:sec:DetectorsDAQ}Detectors and Data Acquisition System}
\begin{figure}
\includegraphics[width=0.475\textwidth]{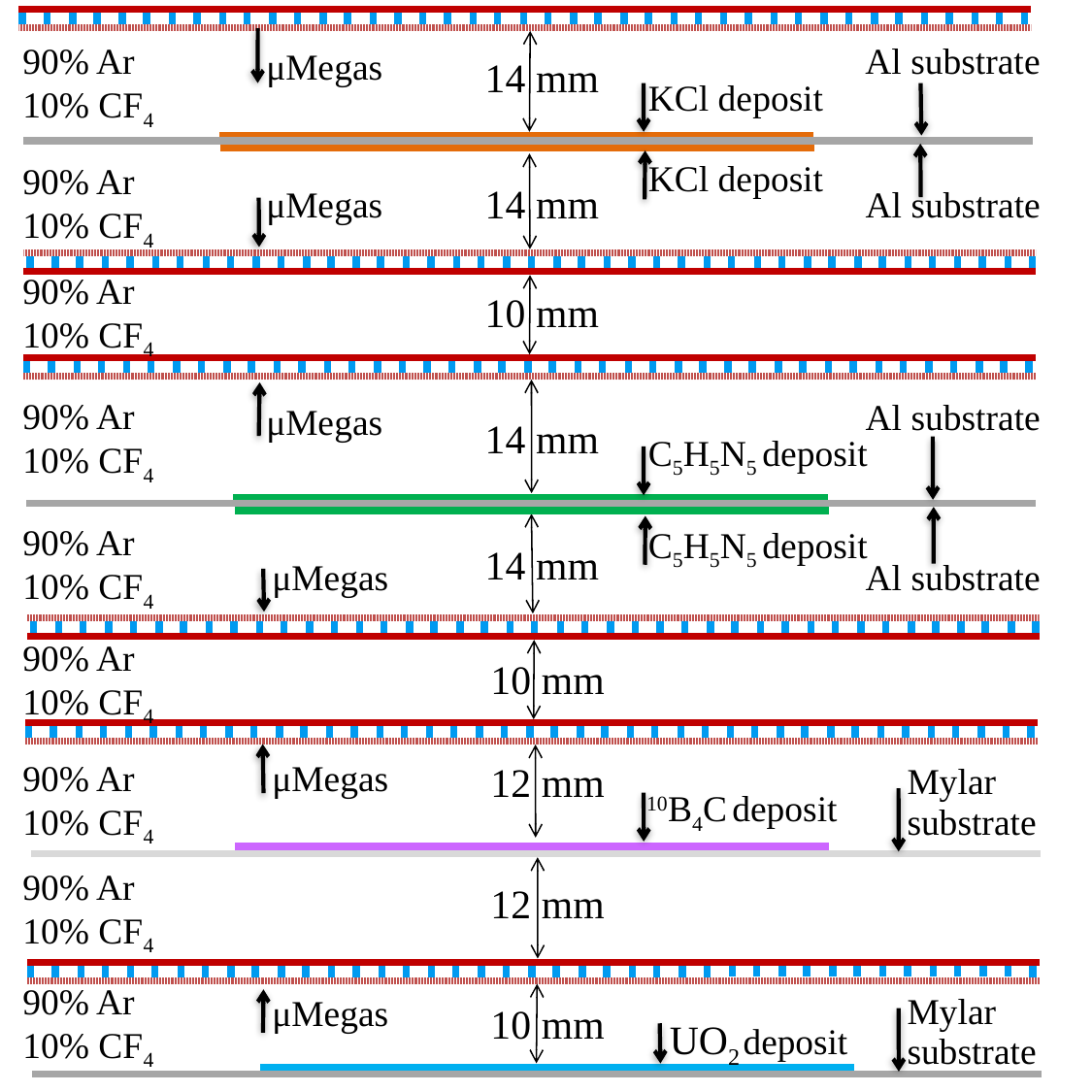}
\caption{\label{fig:representation_setup_micromegas} Schematic representation and dimensions of the microMEGAS detectors in the chamber. The setup consisted of six different samples with a dedicated microMEGAS detector associated to each one. Detectors, deposit thicknesses and substrates were chosen to ensure that all emitted charged particles left their energy in the regions of interest with minimum probability to produce a signal in a detector not associated with the sample.}
\end{figure}

The experimental setup was based on a set of micromesh gaseous structure detectors (microMEGAS). The microMEGAS detectors are based on microbulk technology, for which the low mass, robustness and transparency to $\gamma$ radiation allow the use of several detectors along the beam with minimal perturbation and attenuation of the beam. If at least one product from the reaction is detectable, a geometrical efficiency close to 50\% is ensured \cite{Andriamonje2011}. This type of detectors has been extensively used in n\_TOF fission and charged-particle emission measurements \cite{Weiss2015, Sabategilarte2017, Praena2018A, TorresSanchez2023}. 

A full description of the experimental setup can be found in Ref. \cite{TorresSanchez2023}. Here only necessary information is provided. Fig. \ref{fig:representation_setup_micromegas} shows a schematic representation of the detectors in the chamber. The chamber was filled with a mixture of 90\% Ar and 10\% CF\textsubscript{4}, containing six samples: UO\textsubscript{2} (enriched in \textsuperscript{235}U), \textsuperscript{10}B\textsubscript{4}C, two C\textsubscript{5}H\textsubscript{5}N\textsubscript{5}, and two KCl,  with a detector adjacent to each of them. U and B samples were placed in the forward direction (with respect to the beam-line direction), while Cl and N samples were placed in back-to-back pairs, one microMEGAS facing backwards and the other one forwards, in order to detect possible anisotropies. Results for N samples have already been published in Ref. \cite{TorresSanchez2023}. 

Additional measurements, substituting the N and Cl samples with dummy ones (only Al foils), were also performed in order to determine the background and reduce systematic uncertainties.

The experimental setup was aligned to the nominal beam position. The real position and spatial profile of the beam were checked by the use of Gafchromic foils. These are radiosensitive films that contain a dye that changes color when exposed to ionizing radiation, providing high resolution of the beam profile distribution. The size of the beam-spot at the position of the first sample is 1.9 cm (FWHM). After irradiation, the foils were processed through digital scanning. The information from the digital scanning was adopted for corrections in the efficiency related to the beam-to-sample intersection. 

The detector signals are acquired by the standard n\_TOF data acquisition system based on SPDevices ADQ412DC-3G cards of 2 GS/s maximum sampling rate, 12 bits resolution, and 175 MB on-board memory \cite{Abbondanno2005}. The special features of these cards ensure the collection of data for a Time of Flight corresponding to neutron energies well below the thermal energy. The signal induced by the prompt $\gamma$-flash generated in the interaction of the 20 GeV/c proton beam with the lead target is used as the reference signal for the Time of Flight determination.

\subsection{\label{sec:sec:Samples_characterization}Samples characterization}
All the samples were manufactured as a deposit onto a substrate. The deposits in the samples cover 9 cm in diameter, enough to cover the entire neutron beam. 

The UO\textsubscript{2} deposit is enriched with \textsuperscript{235}U to 99.934\% and prepared using the electrodeposition method. Its thickness is $(123.2 \pm 0.5)$ nm and its areal density $(3.047 \pm 0.013)\cdot 10^{17}$ atoms/cm\textsuperscript{2} onto 30 $\mu$m of Al. 

The \textsuperscript{10}B\textsubscript{4}C (enrichment $>$99.9\%) deposit is prepared by the sputtering method. The \textsuperscript{10}B\textsubscript{4}C areal density is $(1.69 \pm 0.09)\cdot 10^{17}$ atoms/cm\textsuperscript{2} onto 100 $\mu$m of Mylar.

The KCl deposits contain natural potassium chloride of purity $>$99.5\%, and were deposited by thermal evaporation onto 100 $\mu$m Al foils. No enrichment process was needed for \textsuperscript{35}Cl: \textsuperscript{35}Cl has an abundance of 75.5\% in natural chlorine and no charged particles are emitted after neutron interaction with \textsuperscript{37}Cl at relevant neutron energies. In addition, $\alpha$ particles from \textsuperscript{39}K(n,$\alpha)$ have higher energies so they can be filtered by applying cuts in the energy deposited inside the detector, $E_{dep}$, besides having a cross-section around 70 times lower; and $\alpha$ particles from \textsuperscript{41}K(n, $\alpha)$ are only emitted at neutron energies above 120\,keV. The KCl samples were characterized via Rutherford backscattering spectrometry (RBS) of He++ at 4.0\,MeV at the Centro Nacional de Aceleradores (Spain), where previous works had shown excellent possibilities for sample characterization \cite{Praena2018B}. The RBS spectra were analyzed using the SIMNRA package \cite{SIMNRA_manual}.

\begin{figure}[H]
\includegraphics[width=0.5\textwidth]{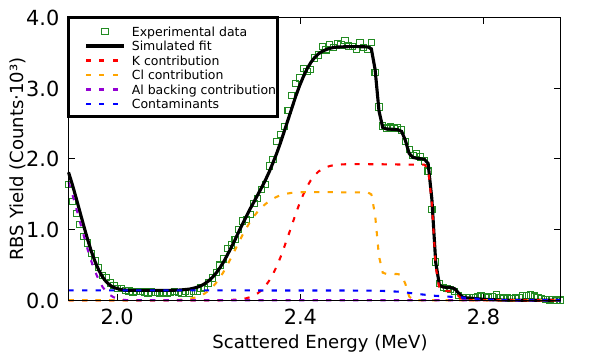}
\caption{\label{fig:simnra_fit} Example of a spectrum from a RBS from one KCl sample. The results of the SIMNRA fit with detailed contributions from each isotope are shown with dashed lines.}
\end{figure}

Given the dimension of the samples (diameter of 9 cm for the KCl deposits and 10 cm for the Al substrates) and of the He++ beam spot (3 mm), several points across the samples were studied to exactly determine the profile of the thickness. The samples were scanned from the edges to the center in three different radial directions. In order to perform an accurate and precise determination of the number of atoms of \textsuperscript{35}Cl, a few points outside the area coated with KCl were also analyzed by RBS. This allowed the determination of any possible contamination present in the Al backings, reducing the free parameters of the SIMNRA fit of the data. Fig. \ref{fig:simnra_fit} shows an example of SIMNRA fit to the data for one sample. 

\begin{figure}[H]
\includegraphics[width=0.5\textwidth]{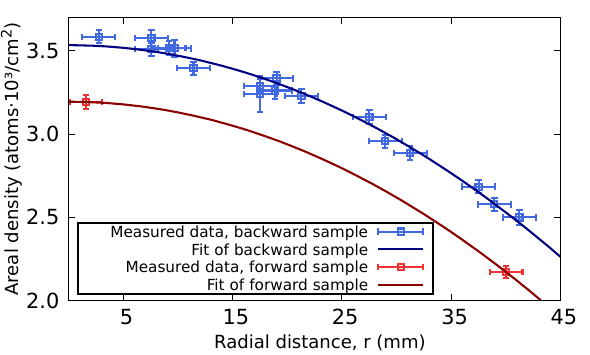}
\caption{\label{fig:masses_1D_fits} Measured distribution and parabolic fits of the deposits areal density as a function of the radial distance $r$ (from the sample centre). Cl forward sample had already been characterized in the past, thus requiring only two measurement points to check whether it had lost or gained mass over time.}
\end{figure}

We find a smooth reduction of the KCl mass density from the center to the edges. Within uncertainties, the same mass density is found for points at the same distance from the center, implying that a radial distribution with no angular dependence is a safe assumption. Figure \ref{fig:masses_1D_fits} reveals a parabolic pattern in the mass density, which is attained through the equation $m(r) = m_0-a\cdot r^2$. Parameter $a$ describes mass distribution throughout the deposit, quantified as the curvature of the quadratic fit, while $m_0$ is the mass density at the center of the deposit.

The total mass is computed as the integral of the mass density given by the formula mentioned above, and an effective areal density can be computed in the same way convoluting the sample thickness profile with the 2D-profile of neutron beam, see Sec. II.D. The effective areal density of each sample is determined with an uncertainty below 1.5\% and is summarized in Table \ref{tab:masses}.

\begin{table}[]
\caption{Characteristics of the deposits: effective areal densities and thickness at center.}
\label{tab:masses}
\begin{tabular}{ccc}
\hline
\hline
\textbf{Sample} & \textbf{Effective areal density} & \textbf{Thickness at center} \\ \hline
Backward               & 219 $\pm$ 2 $\mu g / cm^2$ & 1.105 $\pm$ 0.010 $\mu m$             \\ \hline
Forward                & 198 $\pm$ 3 $\mu g / cm^2$ & 0.998 $\pm$ 0.015 $\mu m$            \\ \hline
\end{tabular}
\end{table}

\subsection{\label{sec:sec:Samples}Attenuation and flux corrections}
Even if the perturbation of the flux introduced by the microMEGAS presence in the beam is small, it is still not negligible. Besides, scattering of neutrons from one sample towards the next one slightly distorts the neutron spectrum at downstream samples. These corrections are taken into account by simulating the experimental setup in the Monte Carlo simulation code MCNP6 \cite{MCNP_user_manual}.

\begin{figure}
\includegraphics[width=0.5\textwidth]{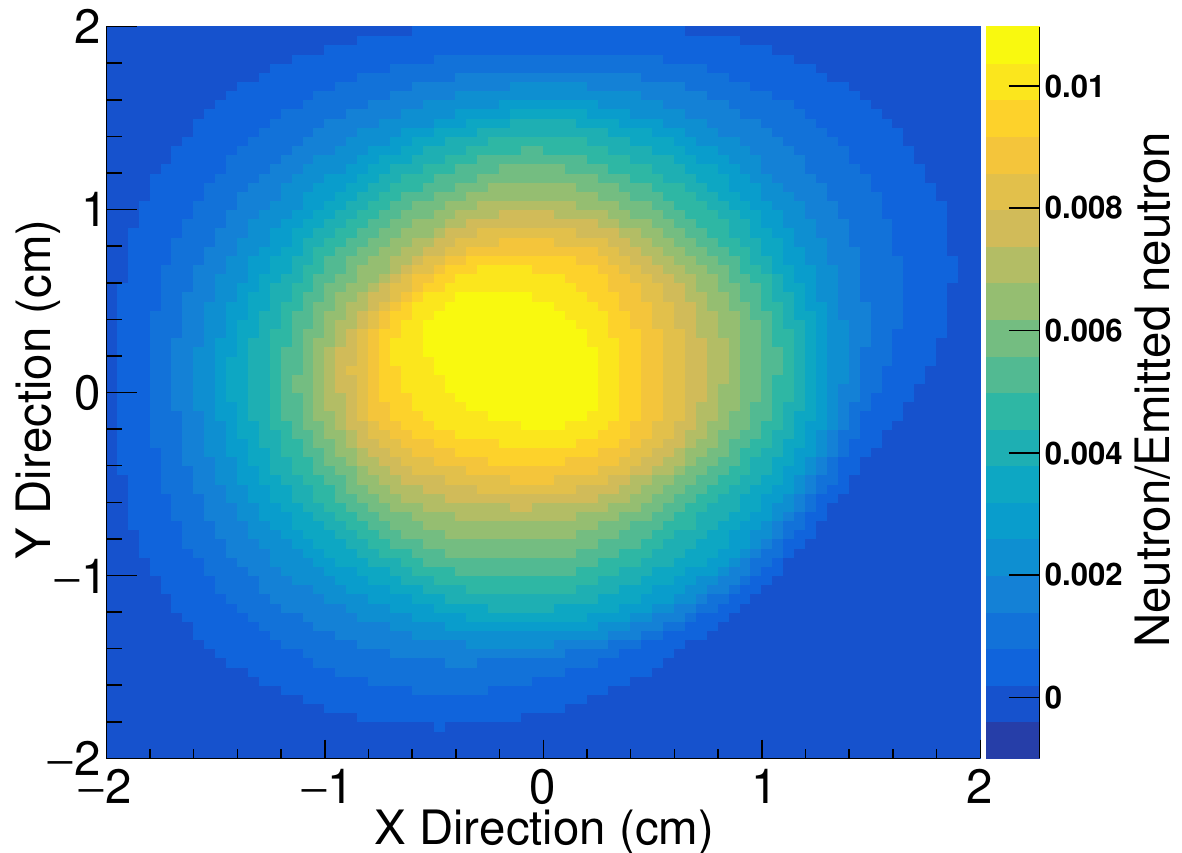}
\caption{\label{fig:profile_2D_neutrons} 2D-profile of the simulated neutron beam at the entrance of the experimental setup (deposit of U). Z-axis is normalized to the total number of emitted neutrons in the simulation done with the Transport Code.}
\end{figure}

\begin{figure}
\includegraphics[width=0.5\textwidth]{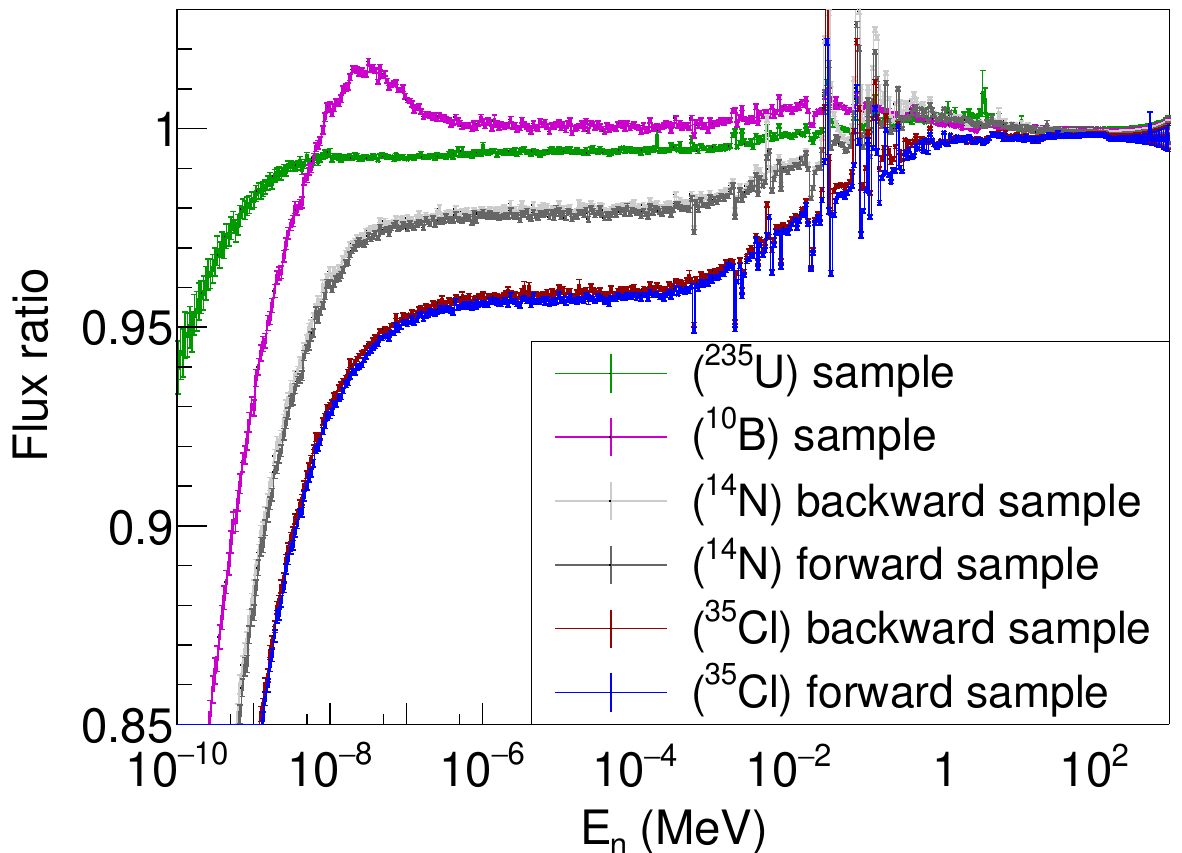}
\caption{\label{fig:transmission_factors} Ratio between the neutron flux at the position of each sample and the incoming neutron flux (evaluated flux from the n\_TOF Transport Code), obtained using Monte Carlo Simulations.}
\end{figure}

The 2D flux profile is well characterized at n\_TOF facility \cite{TC_GUI_user_manual}, and it is shown in Fig. \ref{fig:profile_2D_neutrons} at the position of U sample. The profile is needed not only for determination of above-discussed intensity corrections, but also for determination of effective areal density of the samples. We found that the simulated flux profile is compatible with the results from analysis of the Gafchromic foils. The change in flux for each sample as a function of the neutron energy $E_{n}$ is shown in Fig. \ref{fig:transmission_factors}. 

\subsection{\label{sec:sec:Samples}Efficiency calculation}
For reactions where at least one of the particles emitted is detectable, an efficiency of around 50\% is ensured for microMEGAS detectors. However, exact efficiency slightly depends on the type of particle itself, the deposit from which it is emitted, the thickness it has to traverse before reaching the detector, and more importantly the energy of the particle. At the thermal point, angular spectrum of emission is isotropic but for more energetic neutrons, produced particles tend to move downstream, which slightly affects the detection efficiency. To accurately determine the efficiency as a function of the incident neutron energy for B and Cl microMEGAS detectors, a series of MCNP simulations are done, in which the angular emission and the energy of the emitted particles are modified accordingly for several neutron energies. As for the efficiency of the microMEGAS associated to the U sample, it is described in \cite{TSINGANIS201458}. 

\section{\label{sec:Data_analysis} Data analysis}
The digitized signals are processed offline by means of a pulse shape analysis routine \cite{PSA_proceeding}, from which information is extracted on the amplitude, area, time, and other features of the signals. The analysis is done separately for high intensity (HI) and low intensity (LI) proton pulses from the Proton Synchrotron accelerator complex. HI pulses (corresponding to $\sim 7\cdot 10^{12}$ protons per bunch) allow larger statistics, but suffer from a very intense $\gamma$- flash. On the other hand, LI pulses (corresponding to $\sim 3.5 \cdot 10^{12}$ protons per bunch) allow a better signal identification for higher neutron energy since their $\gamma$-flashes are less intense.

\subsection{\label{sec:sec:TOFtoEnergy_conversion}TOF-to-Energy conversion and mass calibration}
The experimental TOF yield for U and B samples is compared to simulations from the n\_TOF Transport Code \cite{TC_GUI_user_manual}, which includes ENDF/B-VIII.1 cross-section evaluation combined with the effect of the n\_TOF EAR-2 Resolution Function (phase 3) \cite{resolution_function_ntof}. This comparison permits not only a precise determination of the flight path but also an exact determination of the boron deposit mass, while uranium mass was precisely calculated by measuring activity from \textsuperscript{235}U. The n\_TOF EAR-2 Resolution Function is dependent on the type of pulses (HI or LI), so the corresponding type is used accordingly. The extracted effective flight path is 19.35 m for the U sample position. This value corresponds to the geometrical distance of the experimental U sample from the surface of the Pb spallation target. This flight path is adjusted for the subsequent samples according to their position inside the detection chamber.

\subsection{\label{sec:sec:Dead-time_corrections}Dead-time corrections}

\begin{figure}[H]
\includegraphics[width=0.5\textwidth]{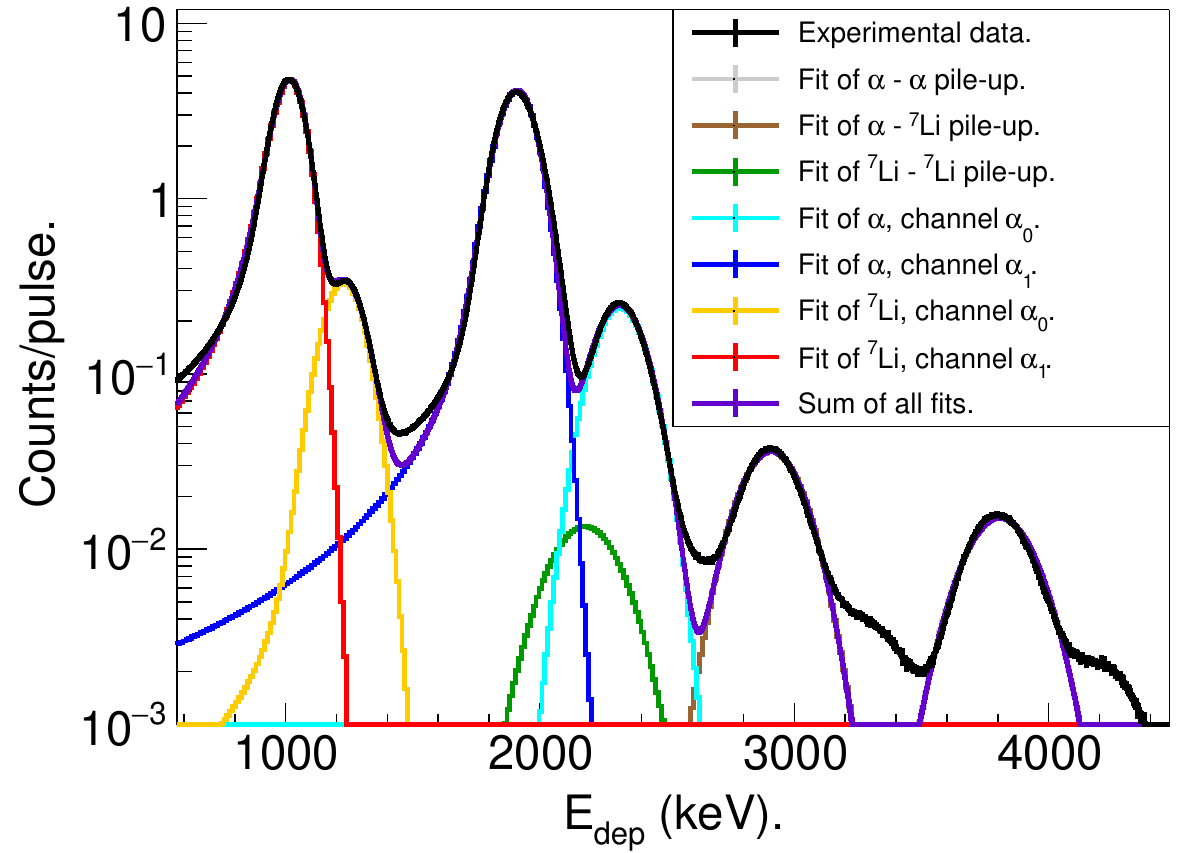}
\caption{\label{fig:dead-time_pileups} Distributions of registered signals in the B microMEGAS as a function of their deposited energy $E_{dep}$ at thermal neutron energy $E_{n}$. Fits of the main possible detection events are also plotted. Neutrons on \textsuperscript{10}B produce $\alpha$ particles and \textsuperscript{7}Li (both detectable) which can be in the ground or in the first excited state (channels $\alpha_0$ and $\alpha_1$, respectively). Fittings of pile-up events are done only with $\alpha$ and \textsuperscript{7}Li particles coming from the $\alpha_1$ channel, since the contribution of pile-ups with particles from $\alpha_0$ channel is notably lower. In the same way, triple (and quadruple, quintuple...) coincidences may also occur but are neglected. Structures formed around $E_{dep} = 3400$\,keV and $E_{dep} = 4200$\,keV, as well as some mismatches between black and purple line (around 600, 1500, 2600\,keV) are explained by those neglected contributions (pile-ups with of $\alpha_0$ channel, and triple coincidences) in conjunction with a background that is also neglected since it is much lower than the main detection peaks. }
\end{figure}

\begin{figure}[H]
\includegraphics[width=0.5\textwidth]{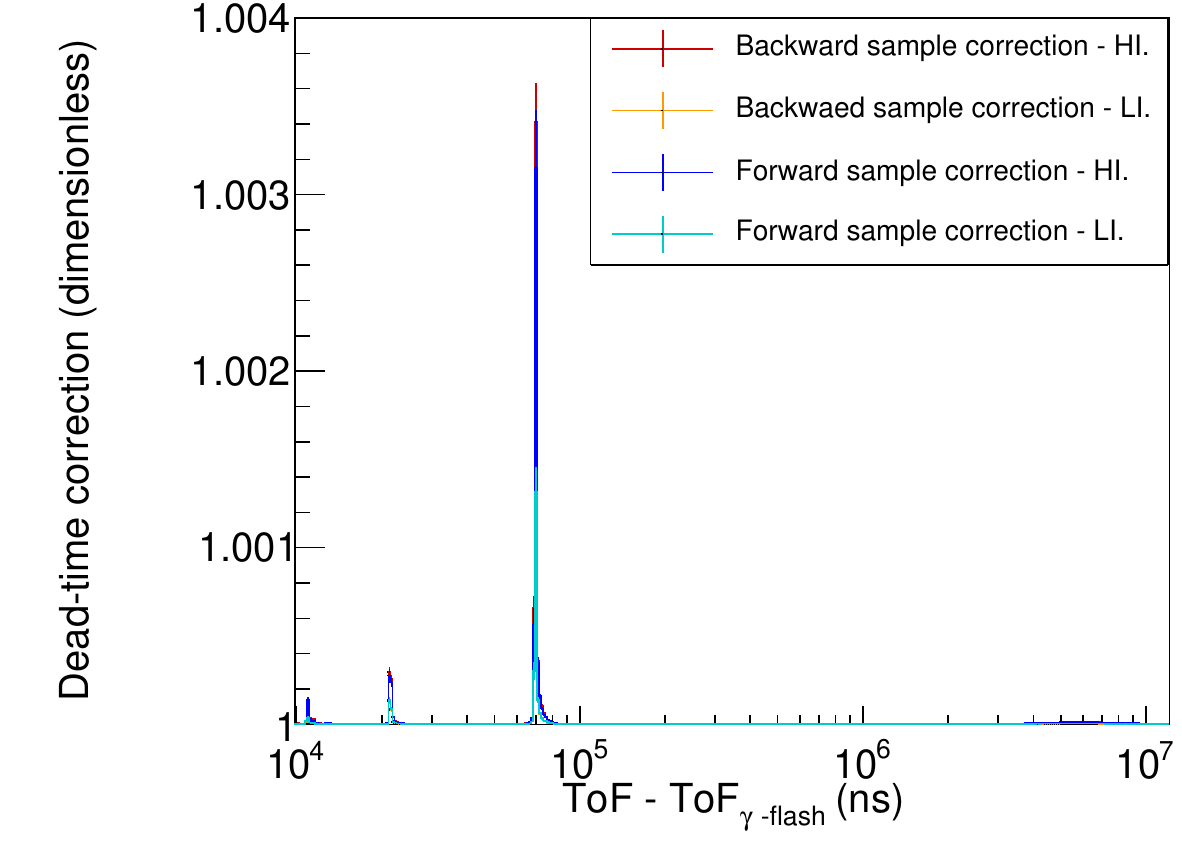}
\caption{\label{fig:dead-time_corrections} Dead-time corrections for both Cl samples (and for HI and LI pulses) as a function of the Time of Flight.}
\end{figure}

Dead-time corrections are determined following the non-paralyzable response model \cite{paralyzable_model} assuming different dead-time for each detector. These dead-times are estimated in two different ways: as the maximum of the Full Width At Half Maximum (FWHM) distribution of all the signals, and as the minimum time difference between consecutive measured pulses. The results from both methods are very similar and the average of them is taken as the estimated dead-time of each detector.

The validity of this correction is verified using two methods. The first one consists on checking that there is a good match between total yields for HI and LI pulses after the correction. The second one is based on checking that the pile-up counts match the dead-time correction. \textsuperscript{7}Li emitted from neutron capture on \textsuperscript{10}B can be either in the ground state (channel $\alpha_0$ with a 6\%) or in the first excited state (channel $\alpha_1$ with a 94\%), creating four peaks in deposited energy inside the detector (see Fig. \ref{fig:dead-time_pileups}). Particles originated in different reactions that enter the detector within its dead time will generate pile-up events. By extracting pile-up contributions and comparing them with those from the dead-time corrections we can verify that they are appropriate. 

Dead-time corrections applied on the TOF histograms are displayed in Fig. \ref{fig:dead-time_corrections}. For the Cl samples the highest value of the dead-time correction $f_{DT, Cl}$ is at the first resonance ($\leq$ 1.004) and the correction is almost negligible (1.00001) around thermal energy, due to a much lower count rate.

\subsection{\label{sec:sec:Background}Background and electronic noise corrections}

\begin{figure}[H]
\includegraphics[width=0.5\textwidth]{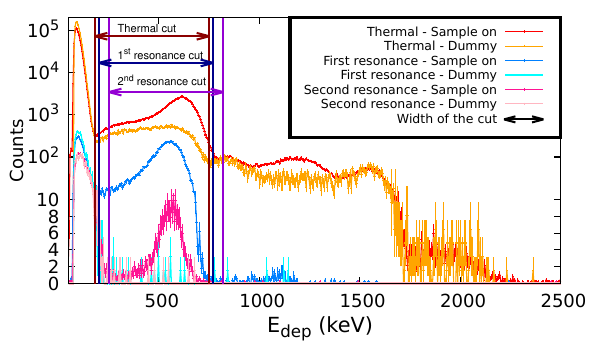}
\caption{\label{fig:background_cuts} Distributions of registered events as function of their deposited energy, at several neutron energy ranges from microMEGAS associated to backward Cl sample (HI pulses). Background has been normalized to the total current of sample-on runs.}
\end{figure}

There is a background in the measured spectra coming from different sources: ambient, related to the backing material and related to other reactions in the sample. Measurements with dummy samples (aluminum backings without KCl deposited) were performed to characterize the ambient and backing background. A contribution from the own sample is present, as shown in Fig. \ref{fig:background_cuts}, as a second peak around deposited energy $E_{dep} \approx$ 1300\,keV,  due to the $\alpha$ particles emitted from \textsuperscript{39}K(n, $\alpha$) reaction, while the protons emitted from $^{35}Cl(n,p)$\textsuperscript{35}S reaction are expected to appear around $E_{dep}\approx$ 600\,keV. These additional events of sample background are removed by a $E_{dep}$ cut, with negligible contribution outside the cut.

Further, there are counts at low deposited energy ($E_{dep}<150-200$\,keV) due to electronic noise. Electronic noise can be filtered by applying another threshold in deposited energy, now at lower values. The cut is not fixed for all neutron energies and becomes larger the closer it is to the $\gamma-$flash, see Fig. \ref{fig:background_cuts}. In practice, some counts from $^{35}Cl(n,p)$ reaction are lost outside the cut, which are estimated by fitting the proton peak to the convolution of an inverse Landau distribution with a gaussian one. The fraction of missing events (due to applied $E_{dep}$ cut) is then characterized by a factor $f_{Cut}$.

\subsection{\label{sec:sec:Uncertainties}Uncertainties}
Different binnings are used at different neutron energy ranges: 500 bins per decade (bpd) below 100\,meV; 500, 250 or 125 bpd for resonances; 50 or 25 bpd between resonances. The statistical uncertainty of the raw spectra ranges from 3\% for HI and 5\% for LI pulses around the thermal point, to 5\% (HI) and 8\% (LI) around resonances. Uncertainties above 20\% are reached in the valleys between resonances, where the cross-section is considerably lower. Uncertainty from background subtraction becomes the most important source of uncertainty in valleys between resonances, due to the fact that the total amount of counts is comparable to the background contribution.

As for systematic uncertainties, conservative estimates are adopted. The mass of the Cl samples is determined with uncertainty below 1.5\%. The uncertainty contribution from the correction of the selection cuts ranges from 2\% at thermal energy to 10\% in the keV range. 

The uncertainty contribution due to the detection efficiency, 1\%, includes the statistical uncertainty in the simulations of the charged particles transport and also, indirectly, the systematic mass uncertainty, since the deposits are simulated as well. The uncertainty contribution due to the presence of the samples (absorption, production of new neutrons in uranium and scattering, represented in Fig. \ref{fig:transmission_factors}) also includes a statistical contribution from Monte Carlo calculations and a systematic one by the mass of the samples included in the simulation. It is less than 1.2\% and is also dominated by the mass uncertainty. Uncertainty in the dead-time correction is very low and neglected in the analysis below.

These contributions are summarized in Table \ref{tab:uncertainties} and total final uncertainty is shown in the lower part of Fig. \ref{fig:yield_Cl}, where several rebinnings have been made to further reduce statistical uncertainty.

\begin{table}[H]
\caption{Major sources of uncertainty (in \%). Statistical uncertainties vary depending on the binning, the energy range and on the sample. MicroMEGAS statistics corresponds to background-substracted spectra and when no specification is given, minimum-maximum values are reported.}
\label{tab:uncertainties}
\begin{tabular}{lr}
\hline
\hline
\multicolumn{1}{l}{Component} & Uncertainty (\%) \\ \hline
MicroMegas statistics           & 4 ($1/v$) - 20 (reson.) - 50 (valleys)            \\
Deposit mass                     & 0.5 - 1.5            \\
Neutron beam attenuation        & 0.6 - 1.2        \\
Efficiency                      & 0.5 - 1          \\
Selection cuts                  & 2 - 10          \\ \hline \hline
\end{tabular}
\end{table}

\section{\label{sec:Results} Results}
\subsection{\label{sec:sec:Cross-section}Cross-section}
\subsubsection{\label{sec:sec:sec:Yield_calculation }Yield calculation}

\begin{figure*}
\includegraphics[width=1.0\textwidth]{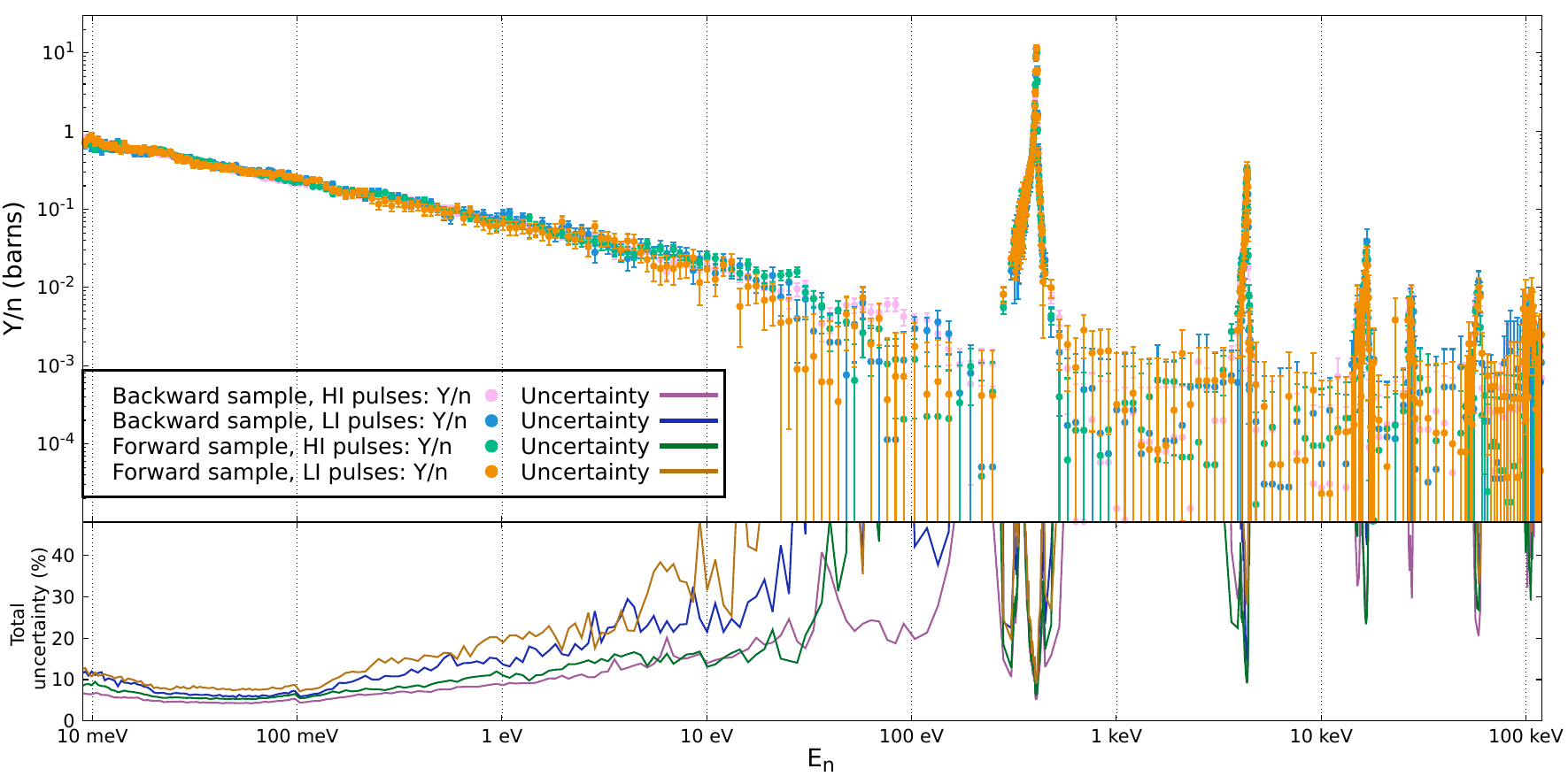}
\caption{\label{fig:yield_Cl} Experimental chlorine yield divided by the sample areal density ($Y/n$). This quantity differs from the cross-section only by the effect of the n\_TOF EAR-2 resolution function. Lower pannel of the plot shows the total uncertainty in percentage at each neutron energy.}
\end{figure*} 

Our measurement covers seven orders of magnitude of neutron energy, providing a common consistent dataset for the thermal, $1/v$ range, and the resonance region. The yield over areal density $Y/n$ is calculated relative to the well-known \textsuperscript{10}B$(n, \alpha)$ cross-section, following 
\begin{widetext}
\begin{equation}
    Y/n (E_n) = \sigma_{^{10}B} (E_n)\cdot\frac{[C_{Cl}(E_n)\cdot f_{DT, Cl}(E_n) - C_{BG}(E_n)\cdot f_{DT, BG}(E_n)]\cdot f_{Cut, Cl}(E_n) \cdot \Phi_{B}(E_n) \cdot n_{B} \cdot \varepsilon_{B}(E_n)}{[C_{B}(E_n)\cdot f_{DT, B}(E_n) - C_{BG}(E_n)\cdot f_{DT, BG}(E_n)]\cdot f_{Cut, B}(E_n) \cdot \Phi_{Cl}(E_n) \cdot n_{Cl} \cdot \varepsilon_{Cl}(E_n)},
\label{eq:yield}
\end{equation}
\end{widetext}

\noindent where $C_{X}$ (X being Cl for the chlorine sample, B for the boron sample or BG for background measurements) is the number of counts inside the chosen $E_{dep}$ cuts; $f_{DT, X}$ is the dead-time correction factor (see Section III.B); $f_{Cut, X}$ is the correction for the energy-cut factor (see Section III.C and Fig. \ref{fig:background_cuts}); $\Phi_{X}$ is neutron flux at sample position; $n_{X}$ is the areal density of atoms; and $\varepsilon_{X}$ is the detection efficiency of the assigned detector.

Relative measurement with respect to B was chosen as calculation of the absolute value of $\Phi_{Cl}$ is not a trivial task and can only be done with a significant uncertainty. By performing a relative measurement to boron, the ratio $\Phi_{B}$ / $\Phi_{Cl}$ in Eq. (\ref{eq:yield}) reduces to ratio of factors shown in Fig. \ref{fig:transmission_factors} and can be determined with a high precision. The boron cross-section is taken from \cite{CARLSON2018143}.

\begin{figure}[H]
\includegraphics[width=0.45\textwidth]{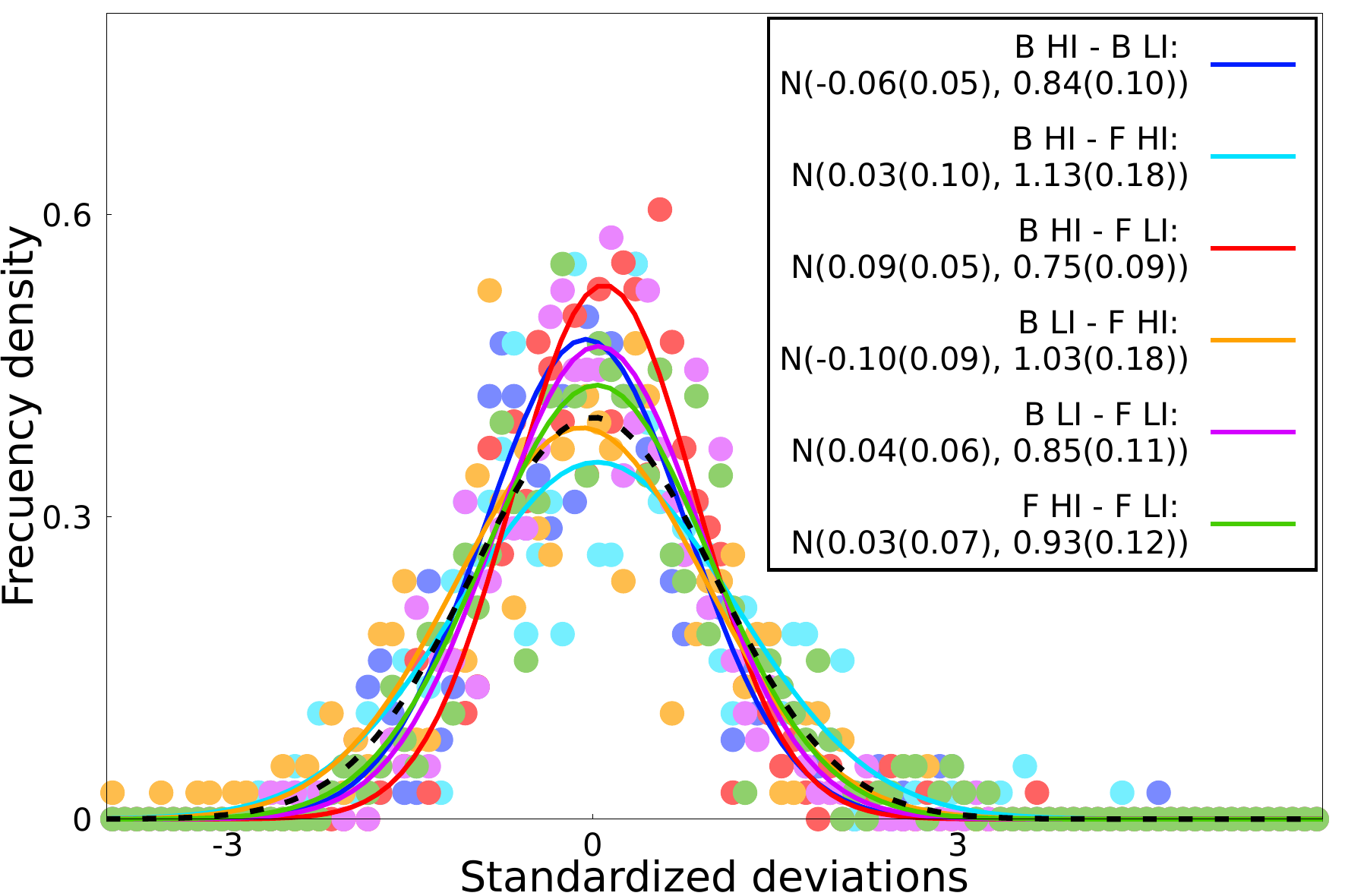}
\caption{\label{fig:yields_fit} Distributions of standardized deviations between different experimental sets. Labels B and F stand for Backward and Forward samples, respectively. Parameters of fits with Normal distribution $N(\mu,\sigma)$.}
\end{figure}

Fig. \ref{fig:yield_Cl} shows the yield (calculated with equation \ref{eq:yield}) divided by sample areal density for the two chlorine samples and LI and HI separately, alongside the relative uncertainty of each point. In order to verify whether the four data sets are statistically compatible and that the correction of systematic effects are well attained, a test is performed and is shown in Fig. \ref{fig:yields_fit}. Distributions of standardized deviations are calculated from two sets $A$ and $B$ as $\eta_{i} = \frac{y^{(A)}_i - y^{(B)}_i}{\sqrt{(\Delta^{(A)}_i)^2 + (\Delta^{(B)}_i)^2}}$ for all points $i$ in the set, where $y_i$ is the value and $\Delta_i$ the total uncertainty of point $i$. As the uncorrelated uncertainty strongly dominates the total uncertainty, individual $\eta_{i}$ are expected to follow a Normal distribution with a zero mean and unit standard deviation $N(0,1)$. As evident from fits in Fig. \ref{fig:yields_fit}, distribution of $\eta_i$ of all possible combinations of sets $A$ and $B$ are nicely consistent with this distribution. This test validates the statistical compatibility of the individual data sets, allowing an uncertainty-weighted-average of the four data sets to be used for the cross-section determination, thus reducing statistical uncertainties.

\subsubsection{\label{sec:sec:sec:Rmatrix_analysis}R-matrix analysis}
\begin{figure*}
\includegraphics[width=1\textwidth]{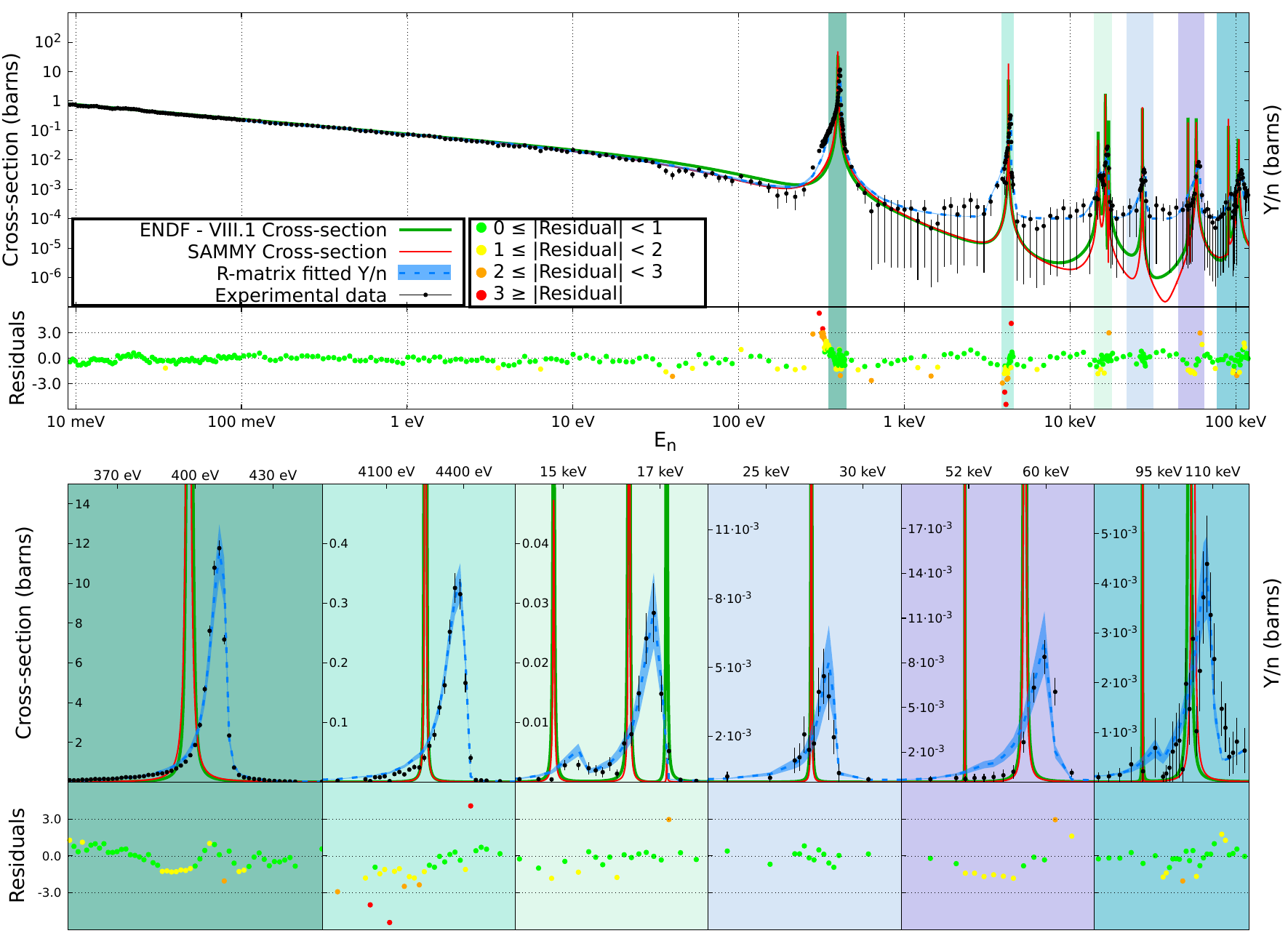}
\caption{\label{fig:xs_SAMMY_all_range} Yield over sample areal density $Y/n$ from experiment (black dots) compared to SAMMY fit (blue area), actual cross-section from our resonance parameters (red) and from ENDF-VIII.1 evaluation (green). Lower panels of the figure zoom to individual resonances. Residuals are then shown as colored dots.}
\end{figure*}

\begin{table*}
\caption{Resonance parameters obtained from fitting and resonance strengths $\omega$. Resonance strengths from other authors and previous evaluations \cite{Druyts1994,Koehler1991,Gledenov1999,ATLAS_Mughabghab,Sayer2006} have been included for comparison. Only values marked with asterisk $^{\ast}$ have been varied.}
\label{tab:parameters}
\begin{tblr}{colspec  = {ccccc|cccccc},
             colsep=2pt,
             row{3} = {color_reso1!50},
             row{4}   = {color_reso2!30},      
             row{5}   = {color_reso345!30},  
             row{6}   = {color_reso345!30},
             row{7}   = {color_reso345!30},
             row{8}   = {color_reso6!30},    
             row{9}   = {color_reso78!30},     
             row{10}   = {color_reso78!30},  
             row{11}   = {color_reso910!50},  
             row{12}   = {color_reso910!50},
             rowsep=1pt,
}
\hline \hline
\SetCell[r=2]{c}{$E$ (eV)} &
\SetCell[r=2]{c}{$J^{\pi}_r$} &
\SetCell[c=3]{c}{$\Gamma$ (meV)} & & &
\SetCell[c=6]{c}{$\omega$ (meV)} \\
\cline{3-5}\cline{6-11}
& & $\Gamma_{\gamma}$ & $\Gamma_{n}$ & $\Gamma_{p}$ &
      This work & Druyts & Koehler & Gledenov & ATLAS & ENDF \\
\hline
$397.36\pm0.20$ & 2\textsuperscript{-} & $1030\pm50^{\ast}$ & $61\pm5^{\ast}$ & $414\pm20^{\ast}$ & $10.5\pm0.9$ & $9\pm1$ & 10 & $11.2\pm2.6$ & $7.2\pm1.2$ & 9.79 \\ \hline
$4250.8\pm1.0$ & 1\textsuperscript{-} & $472\pm25$ & $630\pm40$ & $212\pm25^{\ast}$ & $38\pm4$ & $42\pm2$ & 35 & $40\pm8$ & $41\pm4$ & 40.72 \\ \hline
$14802.0\pm1.0$ & 2\textsuperscript{+} & $346\pm24$ & $32600\pm2300$ & $13.5\pm2.5^{\ast}$ & $8.3\pm1.5$ & $18\pm5$ &  &  & $12\pm5$ & 17.30 \\ \hline
$16356.1\pm1.0$ & 3\textsuperscript{-} & $387\pm15$ & $5980\pm740$ & $80\pm16^{\ast}$ & $65\pm13$ & $131\pm16$ & 64 &  & $112\pm14$ & 131.4 \\ \hline
$17134.0\pm1.0$ & 3\textsuperscript{-} & $800\pm30$ & $14100\pm1300$ & $0.45\pm0.20^{\ast}$ & $0.38\pm0.15$ & $26\pm9$ &  &  & $24\pm8$ & 26.43 \\ \hline
$27346.0\pm1.0$ & 2\textsuperscript{-} & $460\pm30$ & $6000\pm1100$ & $70\pm18^{\ast}$ & $40\pm10$ & $82\pm22$ & 69 &  & $73\pm20$ & 83.61 \\ \hline
$51608\pm5$ & 3\textsuperscript{-} & $45\pm9$ & $2400\pm600$ & $10\pm5^{\ast}$ & $9\pm4$ & $80\pm40$ &  &  & $70\pm40$ & 79.38 \\ \hline
$57812\pm5$ & 2\textsuperscript{-} & $540\pm80$ & $(107\pm14)\cdot10^{3}$ & $700\pm140^{\ast}$ & $430\pm80$ & $620\pm120$ & 860 &  & $560\pm110$ & 614.98 \\ \hline
$90526\pm22$ & 2\textsuperscript{-} & $130\pm70$ & $4200\pm2000$ & $60\pm19^{\ast}$ & $35\pm11$ & $160\pm70$ &  &  & $70\pm30$ & 163.8 \\ \hline
$103520\pm10$ & 1\textsuperscript{-} & $390\pm130$ & $(382\pm57)\cdot10^{3}$ & $2000\pm400^{\ast}$ & $740\pm130$ & $700\pm300$ &  &  & $700\pm300$ & 735.3 \\ \hline \hline
\end{tblr}
\end{table*}

The resonance parameters are obtained from fitting the yields divided by the sample areal density $Y/n$ obtained from Eq. (\ref{eq:yield}) with the R-matrix based SAMMY code~\cite{SAMMY_guide}, applying the Reich-Moore approximation. SAMMY is capable of introducing multiple interaction and self-shielding effects but those are negligible in our experiment given the small thickness of the deposits. 

Fig. \ref{fig:xs_SAMMY_all_range} shows the experimental and fitted $Y/n$, the cross-section calculated by SAMMY after optimization of the resonance parameters, and ENDF/B-VIII.1\cite{Sayer2006} cross-section. The bottom part of the figure shows residuals between experimental and fitted yields. The yield over sample areal density from SAMMY corresponds to temperature of 300 K. A good agreement is reached between fitted (blue area) and experimental (black points) values, specially at $1/v$ region where uncertainties are smaller. A bound state was included in the analysis in order to reproduce the $1/v$ behavior, initial values were taken from \cite{ATLAS_Mughabghab} and fitted. 

The results of the analysis of individual resonances are discussed in the following section. The cross-section is very low between resonances so the background matches the counting rate of sample-on spectrum between isolated resonances, which explains the larger uncertainty bars in Fig. \ref{fig:yield_Cl}. It shows up that a small residual background remains (more notable especially at higher energies), since a low count rate during sample-out measurements prevents its precise determination. This small background is expected to have a smooth energy dependence and is assessed with the SAMMY code fitting. This background, however, has little impact on resonance parameters since the cross-section at resonances is generally one order of magnitude greater than such residual background. 

\begin{figure}
\includegraphics[width=0.485\textwidth]{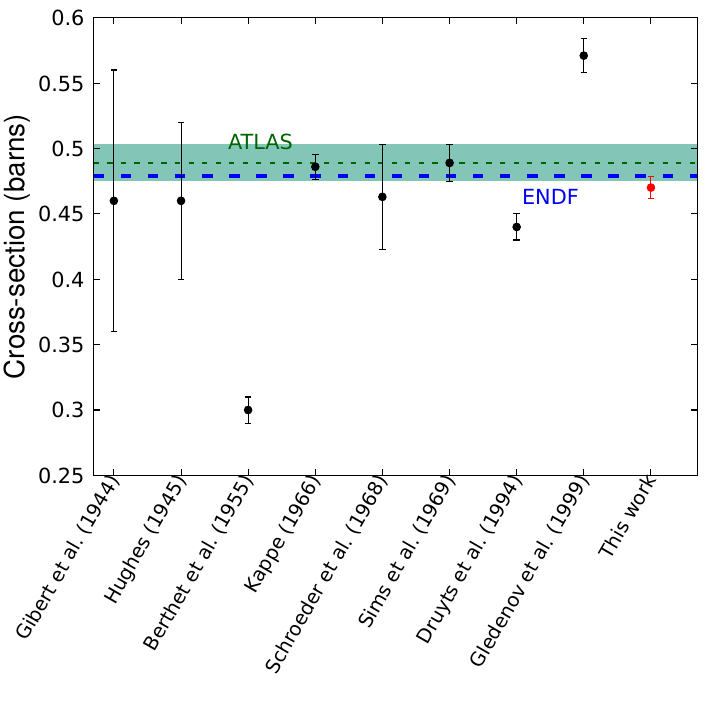}
\caption{\label{fig:thermal_comparison} Comparison of our thermal cross-section with previous measurements and evaluations. As for evaluations, only ENDF \cite{Sayer2006} is considered since the majority of the rest coincide with it. Mughabghab recommended value is also represented with a band of one standard deviation uncertainty. Berthet value is re-normalized to a modern $^{14}N(n,p)$ value.}
\end{figure}

At thermal energy, the obtained cross-section is $0.470 \pm 0.009$ barns. Fig. \ref{fig:thermal_comparison} compares our value with results from previous measurements and values available in evaluations. Our measurement coincides within one standard deviation with all of them except for Druyts \textit{et al.}, Berthet and Gledenov \textit{et al.}. Value from Berthet is remarkably distant. Berthet used \textsuperscript{14}N as reference, whose cross-section was not precisely known in 1955, as posterior modifications prove \cite{TorresSanchez2023}, but even after re-normalization to 1.809 barns (value given in \cite{TorresSanchez2023}) the result is several standard deviations away. Gledenov \textit{et al.} is the most recent measurement but still differs by 22\% from our data. Problems with determination of their sample masses in conjunction with contamination from the \textsuperscript{14}N(n,p)\textsuperscript{14}C reaction, occurring in residual nitrogen present in their ionization chamber could explain this difference.

\subsection{\label{sec:sec:Resonance_analysis} Resonance analysis}

In the lower part of Fig. \ref{fig:xs_SAMMY_all_range}, a zoom to the measured and fitted resonances is included. The region beyond 120\,keV in our measurements is not only very affected by the $\gamma-$flash effects, but also suffers from low statistics, making it impossible to further extend the analysis. For the same reason, higher energy resonances require a wider binning in practice (as mentioned above, 500 bpd for resonances at 0.397 and 4.250\,keV, 250 bpd for the following ones below 30\,keV, and 125 bpd for those at higher energies).

The resonance parameters of observed resonances are given in table \ref{tab:parameters} (fitted parameters for the bound state are $E_R=-100\pm16$\,eV, $J^{\pi}_r=$(2\textsuperscript{+}), $\Gamma_{\gamma} = 0.55\pm 0.11$\,eV, $\Gamma_{n} = 6.3\pm 1.3$\,eV, $\Gamma_{p} = 0.0027\pm 0.0005$ eV). The initial values of all the parameters (energy, spin, parity and partial widths) used in SAMMY fittings were taken from \cite{Sayer2006}. Parameters marked with an asterisk have then been fitted, while all others fixed. The uncertainties in Table \ref{tab:parameters} for fixed parameters are taken from the same evaluation Ref. \cite{Sayer2006} and for fitted parameters from SAMMY. Strengths from previous (n,p) measurements are also included. 

The resonance energies reported in the evaluation \cite{Sayer2006} suited well our measurements in general. As for the partial widths, only the proton width has been varied for every resonance except for the first one. Different values of $\Gamma_{n}$ and $\Gamma_{\gamma}$ affect our results, but in practice the value of $\Gamma_{p}$ dominates the fitting of the experimental yield. Ideally, all parameters should be varied in all resonances, but as statistics get scarcer at higher energies, we have opted to rely on previous dedicated measurements of the other channels and only vary the proton channel. 

Let us note that the most significant deviation of experimental data from the fit is visible at the low-energy tail of the 397\,eV resonance, near 340\,eV. This results from a rather poor knowledge of the Resolution function in this region,  consequence of notable reduction of the flux at this energy due to presence of Mn in the shielding of the spallation target, which increases the uncertainty in the Resolution Function and propagates to the yield reconstruction,. In any case, the impact of this uncertainty itself on the fitted parameters is very low.

Resonance strengths from table \ref{tab:parameters} are calculated as $\omega = g\frac{\Gamma_n\cdot\Gamma_p}{\Gamma_\gamma + \Gamma_n + \Gamma_p}$, where $g$ is the spin factor, $g = \frac{2J_r + 1}{2\cdot(2J_i + 1)}$, $J_r$ is the spin of the resonance (second column of Table \ref{tab:parameters}), and $J_i = \frac{3}{2}$ is the spin of (the ground state of) \textsuperscript{35}Cl. Values from other works are also included in table \ref{tab:parameters}. 

For the first resonance, correlations between partial widths have also been analyzed in determining the resonance strength. The quality of the fit, particularly around the tails of the resonance, is improved by varying all parameters, compared to only varying the proton channel.

For the second resonance, the partial widths are also compatible with all authors and so is its strength. As for the following resonances, lower strengths are systematically found with respect to previous measurements. In particular, notably lower values are found for resonances at 17\,keV and 51\,keV. Values of $0.45\pm 0.20$\,meV and $10 \pm 5$\,meV are given for the $\Gamma_{p}$ of those resonances, which means that the strength of these resonances is at the level of two to three standard deviations consistent with zero and the resonances are barely observed. This suggests that their strengths are indeed much lower than those estimated in \cite{Sayer2006}. For the last two resonances, only Druyts \textit{et al.} (and \cite{Sayer2006}, based on the results from Druyts \textit{et al.}) had previously estimated partial widths, and a lower value is found for the resonance at 90.526\,keV while the last resonance is well compatible with ENDF and Druyts \textit{et al.}

No sign of any additional resonance, aside from those listed in Table \ref{tab:parameters}, was observed in the experimental data.

\subsection{\label{sec:sec:MACS} Maxwellian averaged cross-section}

\begin{figure}
\includegraphics[width=0.50\textwidth]{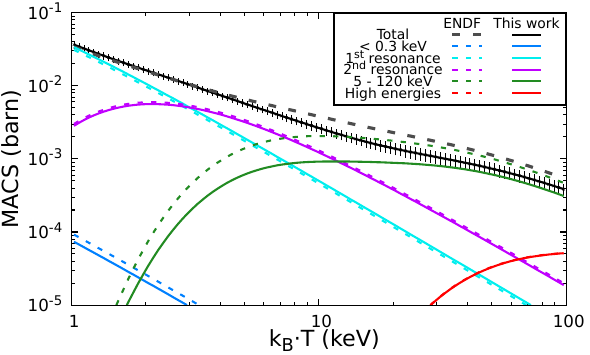}
\caption{\label{fig:MACS} Comparison of Maxwellian Averaged Cross-Sections using the cross-section obtained in this work and ENDF/B-VIII.1 \cite{Sayer2006} evaluation. The ENDF cross-section was used in both MACS for $E_n>120$\,keV. The studied energy range has been divided in five regions: (i) $E_n < 0.3$\,keV, (ii) $0.3$\,keV $\le E_n < 1.0$\,keV (first observed resonance), (iii) $1$\,keV $\le E_n < 5$\,keV (second observed resonance), (iv) $5$\,keV $\le E_n < 120$\,keV (the rest of the observed resonances), and (v) $120$\,keV $\le E_n$ (high energies).}
\end{figure}

The Maxwellian averaged cross-section (MACS) for $^{35}Cl(n,p)$ is calculated in the range of temperatures of astrophysical interest, from $k_B T = 1$\,keV to 100\,keV using (i) the n\_TOF cross-section presented in the previous section and (ii) using ENDF/B-VIII.1 evaluation, see Tab. \ref{tab:MACS} and Fig. \ref{fig:MACS}. The ENDF cross-section for $E_n>120$\,keV was adopted also in (i). Fig. \ref{fig:MACS} illustrates the contribution of different resonances (energy regions) to MACS.  At low temperatures ($k_B T < 5$\,keV), where the first and second resonances contribute the most, agreement within one standard deviation is found between our data and ENDF. The discrepancy grows as the rest of resonances start to contribute more.

\begin{table}[]
\caption{MACS calculated from our results (this work) and using ENDF evaluation.}
\label{tab:MACS}
\begin{tabular}{ccc}
\hline
\hline
\multicolumn{1}{l}{$k_B\cdot$ T (keV)} & This work (mb) & ENDF (mb) \\ \hline
1                  & 3.6 $\pm$ 0.4 & 3.4   \\
5                  & 5.6 $\pm$ 0.6 & 6.4   \\
8                  & 3.3 $\pm$ 0.3 & 4.4    \\
10                  & 2.6 $\pm$ 0.3 & 3.7   \\
15                  & 1.8 $\pm$ 0.3 & 2.8   \\
20                  & 1.38 $\pm$ 0.24 &  2.2    \\
25                  & 1.21 $\pm$ 0.22 & 1.9    \\
30                  & 1.07 $\pm$ 0.20 & 1.7   \\
40                  & 0.86 $\pm$ 0.17 & 1.3    \\
50                  & 0.72 $\pm$ 0.15 & 1.1    \\
60                  & 0.61 $\pm$ 0.12 & 0.93    \\
80                  & 0.47 $\pm$ 0.09 & 0.69    \\
100                  & 0.38 $\pm$ 0.07 & 0.54    \\ \hline \hline
\end{tabular}
\end{table}

\section{\label{sec:Conclusions} Conclusions}
A new measurement of the \textsuperscript{35}Cl(n, p) \textsuperscript{35}S reaction was performed at the EAR-2 of the n\_TOF facility at CERN. The measurement provides data from subthermal to the resonance region for the first time, covering the energy range from 8\,meV to 120\,keV. The cross-section is obtained via fitting of experimental yields using the SAMMY code. 

The obtained thermal cross-section is $(0.470 \pm 0.009)$ barns, in good agreement with ENDF/B-VIII.1 evaluation and many previous measurements. Significant deviation is found only when comparing to Berthet \textit{et al.} and Gledenov \textit{et al.} Our resonance strengths for the first two resonances are well consistent with the values previously reported, while usually smaller values are found for resonances at higher energies. Specifically, the strength adopted by ENDF/B-VIII.1 are for the majority of resonances larger by more than three standard deviations. This propagates into the MACS calculation done with our cross-section data, which shows a lower value at stellar temperatures from $k_B T\approx 5$ to $k_B T\approx 100$\,keV than predictions using ENDF evaluation.

Nuclear data from this work will be submitted to the EXFOR repository. All these data could promote new evaluations of the cross-section of the \textsuperscript{35}Cl(n, p) reaction.

\section{\label{sec:Acknowledgements} Acknowledgements}
This work has been carried out within the framework of project PID2020.117969RB.I00 funded by MICIU/AEI /10.13039/5011000110 33. This work was partially supported by Spanish projects Junta de Andalucía (FEDER Andalucia 2014-2020) P20-00665 and B-FQM-156-UGR20, the Scientific Foundation of the Asociación Española Contra el Cáncer (INNOV223579PORR), La Caixa Foundation (CC21-10047), Health Institute Carlos III (DTS22-00147) and the funding agencies of the n\_TOF participating institutes. This project has received funding from the European Union's Horizon Europe Research and Innovation programme under Grant Agreement No 101057511 (EURO-LABS project). M.M. acknowledges support from the Spanish Ministry of Science, Innovation and Universities under the FPU Grant No. FPU21/02919. Authors thank Mr. Wilhelmus Vollenberg for the preparation of the samples.

\end{document}